\documentclass{article}
\usepackage{mlspconf}
\usepackage{cite}
\usepackage{amsmath,amssymb,amsfonts}
\usepackage{algorithmic}
\usepackage{graphicx}
\usepackage{textcomp}
\usepackage{xcolor}
\usepackage{todonotes}
\graphicspath{{./Figures/}}
\usepackage{caption}
\usepackage{subcaption}
\usepackage[hyphens]{url}
\usepackage{hyperref}
\usepackage[hyphenbreaks]{breakurl}
\title{L3DAS21 CHALLENGE: MACHINE LEARNING FOR 3D AUDIO SIGNAL PROCESSING}
\makeatletter
\def\@name{ \emph{Eric~Guizzo}, \emph{Riccardo F. Gramaccioni}, \emph{Saeid~Jamili}, \emph{Christian~Marinoni}, \emph{Edoardo~Massaro}, \\\emph{Claudia~Medaglia}, \emph{Giuseppe~Nachira}, \emph{Leonardo~Nucciarelli}, \emph{Ludovica~Paglialunga}, \emph{Marco~Pennese}, \\\emph{Sveva~Pepe}, \emph{Enrico~Rocchi}, \emph{Aurelio~Uncini}, \emph{and~Danilo~Comminiello}\thanks{Corresponding author's email: \href{mailto:danilo.comminiello@uniroma1.it}{danilo.comminiello@uniroma1.it}. This work has been supported by ``Progetti di Ricerca Grandi'' of Sapienza University of Rome under grant number RG11916B88E1942F.}\vspace{1em}}
\makeatother
\address{Department of Information Engineering, Electronics and Telecommunications \\ Sapienza University of Rome, Italy}
\begin{document}
%
%
\maketitle
%
%
%
\begin{abstract}
The L3DAS21 Challenge\footnote{\url{www.l3das.com/mlsp2021}} is aimed at encouraging and fostering collaborative research on machine learning for 3D audio signal processing, with particular focus on 3D speech enhancement (SE) and 3D sound localization and detection (SELD). 
Alongside with the challenge, we release the L3DAS21 dataset, a 65 hours 3D audio corpus, accompanied with a Python API that facilitates the data usage and results submission stage.
Usually, machine learning approaches to 3D audio tasks are based on single-perspective Ambisonics recordings or on arrays of single-capsule microphones. 
We propose, instead, a novel multichannel audio configuration based multiple-source and multiple-perspective Ambisonics recordings, performed with an array of two first-order Ambisonics microphones. To the best of our knowledge, it is the first time that a dual-mic Ambisonics configuration is used for these tasks. 
We provide baseline models and results for both tasks, obtained with state-of-the-art architectures: FaSNet for SE and SELDnet for SELD. 
%

This report is aimed at providing all needed information to participate in the L3DAS21 Challenge, illustrating the details of the L3DAS21 dataset, the challenge tasks and the baseline models.

\end{abstract}
%
%
%
%
%
%
\section{Introduction}
\label{sec:intro}
3D audio is gaining increasing interest in the machine learning community in recent years. 
This field of application is incredibly wide and ranges from virtual and real conferencing to game development, music production, autonomous driving, surveillance and many more. 
Tasks like sound source localization, speech and emotion recognition, sound source separation, speech enhancement and denoising, and acoustic echo cancellation, among others, potentially benefit from tridimensional representations of sound field, as they carry additional spatial information \cite{abesser2020review, DBLP:conf/ijcnn/AdavannePV18}.
3D audio formats permit to obtain an impressive performance in many machine learning-based tasks, usually bringing out a significant improvement over the single/dual-channel formats \cite{DBLP:conf/ijcnn/AdavannePV18, DBLP:conf/icassp/AdavannePV17}. In this context, Ambisonics prevails among other 3D audio formats for its simplicity, effectiveness and flexibility. 

The L3DAS21 Challenge organized within the L3DAS (Learning 3D Audio Sources) project\footnote{Further information on the L3DAS project can be found on \url{www.l3das.com}.} is designed to encourage and foster research on machine learning for 3D audio signal processing. In particular, we focus on two 3D audio tasks: 3D Speech Enhancement (SE) and 3D Sound Event Localization and Detection (SELD), both relying on multiple-source and multiple-perspective (MSMP) Ambisonics recordings.

3D SE aims at removing unwanted information from spurious spatial vocal recordings and further enhancing the speech intelligibility and clarity.
A widespread strategy to perform SE is to use deep neural networks (DNNs) to estimate a time-frequency mask in the Fourier domain that extracts clean speech signals from noisy spectra \cite{DBLP:journals/taslp/WangC18a}.
Neural beamforming techniques as Filter and Sum Networks (FaSNet) \cite{DBLP:conf/asru/LuoHMCL19} provide state-of-the art results for Ambisonics-based SE and are usually suitable for low-latency scenarios. 
Also U-Net-based approaches provide competitive results in this context, both for monaural  \cite{DBLP:journals/eswa/GuimaraesNS20, DBLP:journals/corr/abs-1811-11307} and multichannel SE tasks \cite{DBLP:conf/eusipco/BoscaGPK20}, at the expense of higher computational power demand.
Other techniques to perform SE include recurrent neural networks (RNNs) \cite{DBLP:conf/icassp/HuangKHS14}, graph-based spectral subtraction \cite{DBLP:journals/speech/YanYWG20}, discriminative learning \cite{DBLP:conf/interspeech/FanLTYW19}, dilated convolutions \cite{DBLP:journals/taslp/LuoM19, DBLP:journals/corr/abs-1905-01697}.

3D SELD, instead, aims at obtaining exhaustive spatiotemporal descriptions of 3D acoustic scenes, predicting which sound categories are present in the scene, and when and where each sound instance is active. 
SELD can be considered as a joining of the traditional sound event detection and sound source localization tasks, and it was presented for the first time in the DCASE2019 Challenge \cite{DBLP:journals/taslp/PolitisMAHV21}.
Also here, the state-of-the-art methods are based on deep learning strategies \cite{DBLP:journals/corr/abs-2006-01919}. SELDnet \cite{DBLP:journals/jstsp/AdavannePNV19} adopted a convolutional-recurrent design with two distinct branches for localization and detection and it was used as a baseline model in SELD tasks of the DCASE challenges.
An improved SELDnet model was then introduced by \cite{DBLP:conf/eusipco/GuirguisSGAY20}, including temporal convolutions.
Other novel solutions for this task include ensemble models \cite{chytas2019hierarchical}, multi-stage training \cite{DBLP:journals/corr/abs-1905-00268} and bespoke augmentation strategies \cite{DBLP:journals/corr/abs-1910-04388, pratik2019sound}.

These tasks are complementary each other and are aimed at fulfilling real-world needs related to real and virtual conferencing. 
Especially in multi-speaker scenarios, it is in fact very important to properly understand the nature of a sound event and its position within the environment, what is the content of the sound signal and how to leverage it at best for a specific application (e.g., teleconferencing and assistive listening or entertainment, among others).

Alongside with the challenge, we present the L3DAS21 datasets, aimed at solving SE and SELD tasks making use of MSMP Ambisonics files, obtained performing 3D audio recordings with an array of two Ambisonics microphones, as further discussed in the next Section.
For the first time, to our best knowledge MSMP dual-mic Ambisonics recordings are considered for machine-learning purposes, giving us the possibility to test the effectiveness of this particular 3D audio format.
Furthermore, the SELD task of the L3DAS21 for the first time proposes a scenario where multiple sounds of the same class may be active at the same time. 
This is a well-established scenario in vision-related object detection tasks and, to our knowledge is an important real-life-oriented study case also in the audio domain.

We supply baseline models and results for both tasks, obtained using state-of-the-art deep learning architectures. Datasets and models are supported by a Python-based API aimed at facilitating the data download and preprocessing, the baseline models training and the results submission.

The L3DAS21 Challenge is open to all and the registration is free of charge. 
All provided materials (datasets, scripts and baseline models) are available for free as well and can be used for any purpose beyond the challenge, under the creative commons CC BY 4.0 license. 

%
%
%
%
%
\section{DATASET DESCRIPTION}
\label{sec:data}
The LEDAS21 dataset contains approximately 65 hours of MSMP B-format Ambisonics audio recordings. 
%
We sampled the acoustic field of a large office room with the approximate dimensions of 6 m (length) by 5 m (width) by 3 m (height). 
The room has typical office furniture: desks, chairs and a wardrobe. 
The floor is made of wood parquet, while the walls and the ceiling are made of painted concrete.

\begin{figure}
 \centering
    \begin{subfigure}[tb]{0.2\textwidth}
    \centering
    \includegraphics[width=4.0cm]{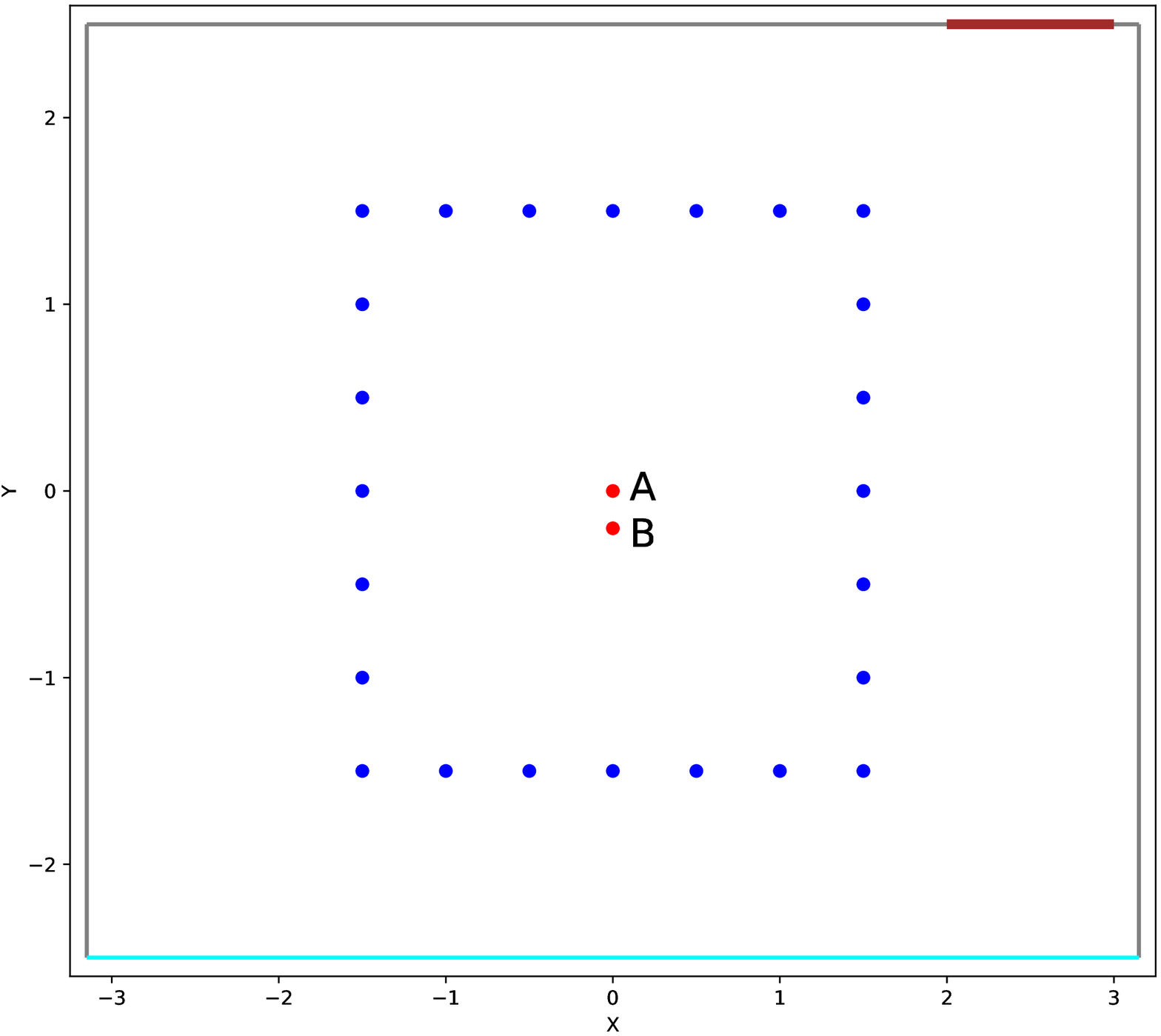}
    \caption{Grid}
    \label{fig:gridpositions}
    \end{subfigure}
    \begin{subfigure}[tb]{0.2\textwidth}
    \centering
    \includegraphics[width=4.0cm]{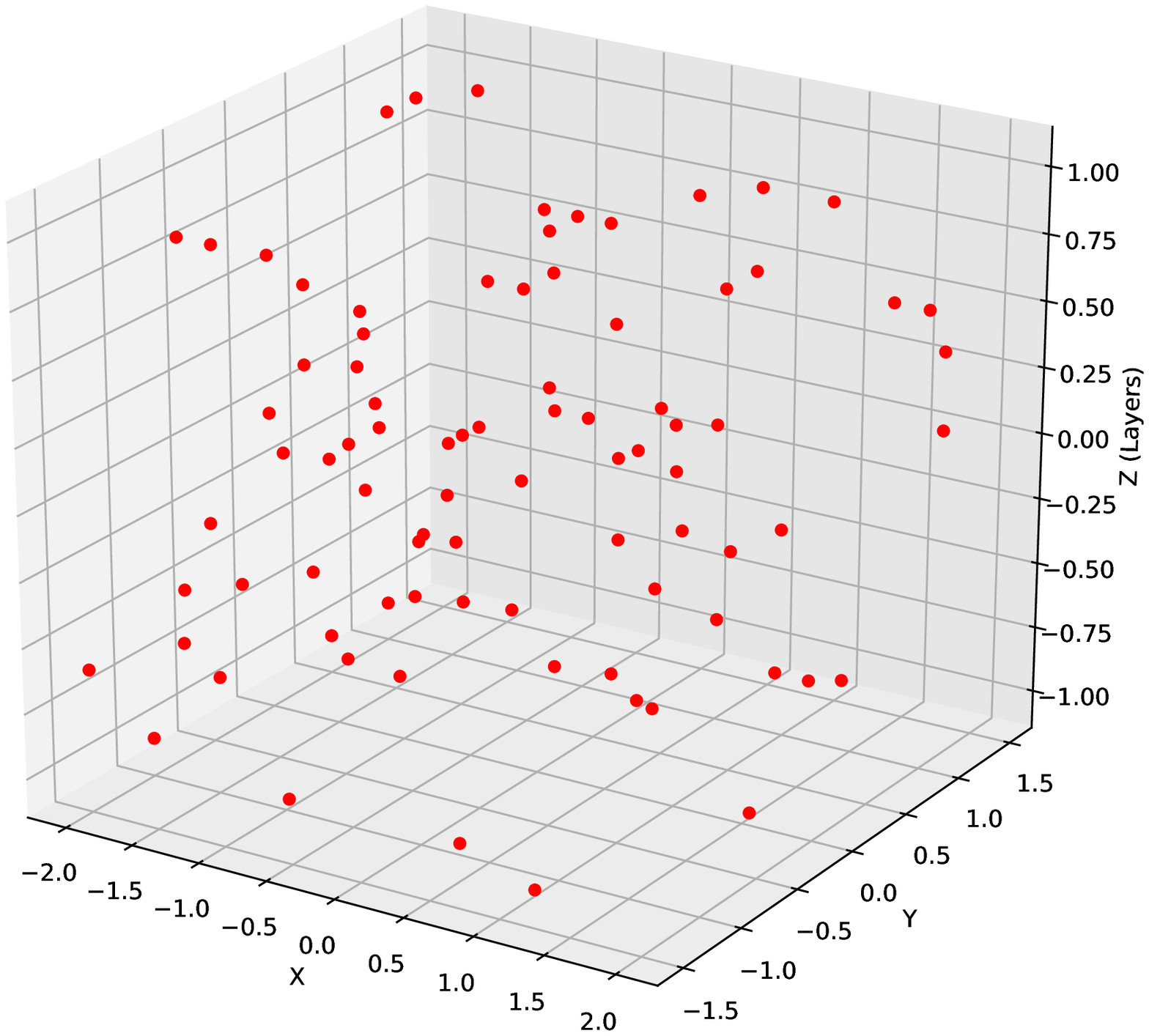}
    \caption{Random}
    \label{fig:randompositions}
    \end{subfigure}
    \caption{(a) Projection from above of the microphones position (A-B red dots) and the speaker positions of the fixed 3D grid (blue dots).
    (b) Tridimensional distribution of the randomly-selected speaker positions.}
    \label{fig:totalpositions}

\end{figure}

We placed two first-order A-format Ambisonics microphones\footnote{Oktava MK-4012} in the center of the room and we moved a speaker\footnote{Event PS6} reproducing an analytic signal in 252 fixed spatial positions.
One microphone (mic A) lies in the exact center of the room, shown as a red dot in Fig.~\ref{fig:gridpositions}, and the other (mic B) is 20 cm distant towards the width dimension.
Both microphones are positioned at the same height of 1.3 m, which is the average ear height ear of a seated person.
The capsules of both mics have the same orientation.

The speaker placement is performed according to two different criteria: a fixed 3D grid (168 positions) and a 3D uniform random distribution (84 positions).
Figure \ref{fig:gridpositions} shows a 2D projection of the grid from above. 
For the first criterion, we placed the speaker in a 3D grid with a 50 cm step in the length-width dimensions and a 30 cm step in the height dimension, as represented in Fig.~\ref{fig:gridpositions} with blue dots.
There are 7 position layers in the height dimension at 0.3 m, 0.7 m, 1 m, 1.3 m, 1.6 m, 1.9 m, 2.3 m from the floor. 
The random positions, instead, respect a uniform distribution and are depicted in Fig.~\ref{fig:randompositions}.
All random-selected positions are quantized in a virtual 3D grid with a 25 cm step.
For all measurements we directed the speaker's tweeter towards mic A.

The analytic signal is a 24-bit exponential sinusoidal sweep that glides from 50 Hz to 16000 Hz in 20 seconds, reproduced at 90 dB SPL on average.
The IR estimation is then obtained by performing a circular convolution between the recorded sound and the time-inverted analytic signal, as introduced by \cite{farina2000simultaneous}.
We finally converted the A-format signals into standard B-format IRs\footnote{\url{http://pcfarina.eng.unipr.it/Public/B-format/A2B-conversion/A2B.htm}}.

Relying on the collected Ambisonics impulse responses, we augmented existing clean monophonic datasets to obtain synthetic tridimensional sound sources by convolving the original sounds with our IRs. 
The result of this convolution operation is the virtual placement of a sound source in the spatial position occupied by the speaker, perceived from the position of the 2 microphones.
We aimed at creating plausible and variegate 3D scenarios to reflect office-like situations, in which disparate types of sound sources and background noises coexist in the same 3D reverberant environment.

For this purpose we used the Librispeech \cite{DBLP:conf/icassp/PanayotovCPK15} and FSD50K \cite{DBLP:journals/corr/abs-2010-00475} 
datasets.
We selected a total of 1440 noise sound files from FSD50K, divided into 14 transient noise classes:\textit{ computer keyboard, drawer open/close, cupboard open/close, finger snapping, keys jangling, knock, laughter, scissors, telephone, writing, chink and clink, printer, female speech, male speech}, and 4 continuous noise classes: \textit{alarm, crackle, mechanical fan, microwave oven}.
We collected 80 sounds for each noise class (both for transient and continuous noises).
Furthermore, we extracted clean speech signals  (without background noise) from Librispeech, taking only sound files up to 10 seconds.

The dataset is divided in two main sections, respectively dedicated to the challenge tasks. 
We provide normalized raw waveforms of all Ambisonics channels (8 signals in total) as predictors data for both sections, the target data varies significantly. 
Moreover, we created different types of acoustic scenarios, optimized for each specific task. 

In addition to this, we include a first-order Ambisonics decoder (supporting decoding to mono, stereo and binaural formats) as part of the supporting API, in order to facilitate the use of our dataset with formats that are easier to handle and can be used also for different applications.

\subsection{SE Dataset}
\label{sec:dataset_se}\

This dataset section is dedicated to task 1 and, thus, is optimized for SE. 
Here we created more than  30000 virtual 3D audio environments with a duration up to 10 seconds each, reaching a total duration of approximately 50 hours.
In each data point a speech signal is always present, mixed with various types of background noise.
We extracted all sounds from the \textit{clean} subset of Librispeech (approximately 53\% male and 47\% female speech).
We add up to 3 non-speech background noises of the above-mentioned categories, extracting them from FSD50K.
With a 25\% chance, one of the background noises is a continuous  noise.
The signal-to-noise ratio ranges from 6 to 16 dB full scale (dBFS), where the voice is always the prominent signal. We randomly place all sound sources in the 3D environment, paying attention to obtain a uniform distribution of locations within this dataset section.

The predictors data of this section are released as 8-channel 16 kHz 16 bit wav files, consisting of 2 sets of first-order Ambisonics recordings.
The channels order is [WA, ZA, YA, XA, WB, ZB, YB, XB], where A/B refers to the used microphone and WXYZ are the B-format ambisonics channels.
The target data provided for this section contains the clean monophonic recordings of the only speech signals (16 kHz 16 bit mono wav files), as well as the words uttered in each data point (in a csv file).

\subsection{SELD Dataset}
\label{sec:dataset_seld}\
This dataset section is targeted to task 2 and it is therefore optimized for SELD.
Here we synthesized 900 1-minute-long data points, reaching a total length of 15 hours of audio. 
Each data point contains a simulated 3D office audio environment in which up to 3 simultaneous acoustic events may be active at the same time.
Moreover, when multiple sounds are active at the same time,  with an approximate probability of 12\% at least 2 sounds may belong to the same class.

The data points of this section contain an average of 26 acoustic events, with a standard deviation of 11.
The sound events belong to the afore-mentioned 14 transient noise classes and are therefore 1120 in total. 
As opposed to the SE dataset, here the data points are not forced to contain speech signals, although they may contain voice sounds. The volume difference between the different sounds ranges from 0 to 20 dBFS.
Also here, we randomly place all sound sources in the 3D environment, paying attention to obtain a uniform distribution of locations.

The predictors data of this section have the same form of the SE section, except the sampling frequency, which here is of 32 kHz.
As target data, we provide a csv file containing the onset and offset time stamps, the typology class and the spatial coordinates of each individual sound event present in a data point.

\subsection{Dataset Splits}
\label{sec:dataset_splits}\
We split both dataset sections into a training set (~44 hours for SE and 600 hours for SELD) and a test set (~6 hours for SE and 5 hours for SELD), paying attention to create similar distributions. 
The train set of the SE section is divided in two partitions: train360 and train100, and contain speech samples extracted from the correspondent partitions of Librispeech (only the sample) up to 10 seconds). 
All sets of the SELD section are divided in: OV1, OV2, OV3. 
These partitions refer to the maximum amount of possible overlapping sounds, which are 1, 2 or 3, respectively. 

The test set of both dataset sections is further split into two equally-long subsets that present a similar distribution: one development and one blind test set. 
The first one is part of the initial release of the dataset, and it is aimed, as usual, at the model's hyperparameters fine-tuning. 
The latter, instead, is released at a second stage and contains the only predictors data, without target labels/signals.
Participants must submit the only results obtained for the blind test set, following the instructions present in the documentation on the GitHub page\footnote{\url{https://github.com/l3das/L3DAS21}}.

\section{CHALLENGE TASKS}
\label{sec:tasks}

We propose 2 different tasks, both based on our L3DAS21 dataset: \textit{3D Speech Enhancement in Office Reverberant Environment} and \textit{3D Sound Event Localization and Detection in Office Reverberant Environment}.
Each one is divided in 2 sub-tasks: one-mic and dual-mic recordings, respectively relying on the sounds acquired by one or both Ambisonics microphones, as described in Section \ref{sec:data}.

In this context, the information predicted for one task may be beneficial for the other one.
For instance, the sound localization parameters may be re-used to improve the performance of 3D speech enhancement networks, as in \cite{DBLP:conf/eusipco/ChazanHHGG19, DBLP:journals/corr/abs-2010-11566}. 
Therefore, participants are encouraged to develop a strategy to bootstrap the resources and exploit the output of one model to enhance the performance of the other one (although this is not mandatory).

%
\subsection{Task 1: 3D Speech Enhancement in Office Reverberant Environment}
\label{subs:speechenh}
The objective of this task is the separation and enhancement of speech signals immersed in a noisy 3D environment, basing on the SE section of the L3DAS21 dataset.
Here the models are expected to extract the monophonic voice signal from the 3D mixture that contains various background noises.  
The evaluation metric for this task is the short-time objective intelligibility (STOI) \cite{DBLP:conf/icassp/TaalHHJ10}, which estimates the intelligibility of the output speech signal. 
Moreover, word error rate (WER) is also computed to assess the effects of the enhancement for speech recognition purposes.
For this purpose, we use a Wav2Vec \cite{DBLP:conf/nips/BaevskiZMA20} architecture pre-trained on Librispeech 960h \footnote{\url{https://huggingface.co/facebook/wav2vec2-base-960h}}.
The final metric for this task is a combination of these two measures given by \((STOI+(1-WER))/2\).
This metric lies therefore in the 0-1 range and higher values are better.

\subsection{Task 2: 3D Sound Event Localization and Detection in Office Reverberant Environment}
\label{subs:seld}
The aim of this task is to detect the temporal activity, spatial position and typology of a known set of sound events immersed in a synthetic 3D acoustic environment.
This task is performed on the SELD section of the L3DAS21 dataset.
Here the models are expected to predict a list of the active sound events and their respective location at regular intervals of 100 milliseconds.

We use a joint metric for localization and detection: Location-sensitive detection error, as defined in \cite{DBLP:conf/waspaa/MesarosAPHV19}. 
This metric is computed on each time frame and consists of measuring the cartesian distance between the predicted and true events with the same label, and counting a true positive only when its label is correct and its location is within a threshold from its reference location. 
After this operation, we compute the regular F score. 
In this challenge, we fixed the spatial error threshold to 2.


%


\section{BASELINE METHODS}
\label{sec:reseval}
As baseline methods we propose state of the art architectures, specifically adapted for each task.

For Task 1 (SE), we use a Filter and Sum Network architecture (FaSNet)  \cite{DBLP:conf/asru/LuoHMCL19}, adapted from this public PyTorch implementation \footnote{\url{https://github.com/yluo42/TAC}}.
This network is a state-of-the-art neural beamformer that operates in the time domain and, therefore, work on both the magnitude and the phase information of the signal.
This baseline model reaches 0.62 for the joint Task 1 metric, with 0.46 WER and 0.72 STOI.

For Task 2, instead, we use a variant of the SELDnet architecture \cite{DBLP:journals/jstsp/AdavannePNV19}.
We ported to the PyTorch language the original Keras implementation \footnote{\url{https://github.com/sharathadavanne/seld-net}} and we modified its structure in order to make it compatible with the L3DAS21 dataset. 
We augmented the capacity of the network by increasing the number of channels and layers, while maintaining the original data flow.
In addition, we the added ability to detect multiple sound sources of the same class that may be active at the same time through an augmented output matrix.
This baseline model reaches an F-score of 0.45 on the location-sensitive detection metric for Task 2, with 0.52 precision and 0.4 recall.

For further details on our baseline models, please refer to the L3DAS official GitHub repository (link above).

%
\section{CHALLENGE RULES AND EVALUATION CRITERIA}
\label{sec:rules}

\subsection{Rules and Requirements}
\label{subs:rr}
The goal of the challenge is to foster research on machine learning for 3D audio. All participants should adhere to the following rules to be eligible for the challenge:
\begin{itemize}
    \item All participants must submit the obtained results for at least one of the 2 tasks, but for both sub-tracks (1 and 2 mics). The results should be accompanied by a paper describing the proposed method.
    \item Each individual participant cannot be included in multiple participating teams. Therefore, a participant is allowed to submit only one set of results.
    \item Winners will be selected according to the best performance for each single task, separately. Therefore, one winner for each task will be selected.
    \item There are no restrictions on the proposed methodologies. However, in case of a tie, the Challenge Committee will take into account the novelty and originality of the proposed approach. Also, a method that can be used for both the 1-mic and 2-mic configurations will be positively evaluated.
    \item Participants are not restricted to use the L3DAS21 dataset only. It is in fact allowed to augment this dataset and/or to integrate additional data to train/pre-train the models.
    \item Accepted papers will be presented at a special session of the IEEE MLSP 2021 on "Machine Learning for 3D Audio Signal Processing". Authors, who are not interested in participating the challenge but want to make a contribution to the topic, are encouraged to submit a paper to this track, even without specifically use the proposed datasets.
\end{itemize}

\subsection{Paper and Results Submissions}
\label{subs:submissions}
%
\begin{itemize}
    \item Results and paper must be submitted within the deadlines shown in Subsection~\ref{subs:timeline} via the IEEE MLSP 2021 submission site\footnote{\url{https://cmt3.research.microsoft.com/MLSP2021}}.
    \item Results should be prepared and formatted according to the guidelines detailed in the challenge submission page\footnote{\url{www.l3das.com/mlsp2021/submission}}.
    \item The accompanying paper must describe the proposed method and must contain all the details to ensure reproducibility. The paper must also include information about the computational complexity of the model (e.g., in terms of number of parameters or execution time on a specific device).
    \item Papers should be prepared according to the guidelines of IEEE MLSP 2021\footnote{\url{https://2021.ieeemlsp.org/paper-submission}}. 
    \item Submitted papers will undergo the standard peer-review process of IEEE MLSP 2021.
\end{itemize}

\noindent Optionally, you can inform us at \href{mailto:l3das@uniroma1.it}{l3das@uniroma1.it} about your submission.

\subsection{Support}
\label{subs:support}
The challenge participants are encouraged to contact our team at \href{mailto:l3das@uniroma1.it}{l3das@uniroma1.it} for any issue of clarification about the challenge or the dataset. 

%
\subsection{Timeline}
\label{subs:timeline}
\begin{itemize}
    \item \textbf{27 Mar 2021} – Release of the datasets (training and development sets).
    \item \textbf{16 Apr 2021} – Release of supporting code, baseline methods and documentation.
    \item \textbf{10 May 2021} – Release of the evaluation test set
    \item \textbf{20 May 2021} – Deadline for submitting results for both tasks
    \item \textbf{27 May 2021} – Notification of the results of participants 
    \item \textbf{31 May 2021} – Deadline for 6-page paper submission
    \item \textbf{31 Jul 2021} – Notification of paper acceptance
    \item \textbf{02 Aug 2021} – Notification of challenge winners 
    \item \textbf{31 Aug 2021} – Deadline for camera-ready papers
    \item \textbf{25 Oct 2021} – Opening of the IEEE Workshop of MLSP 2021
\end{itemize}

%
%

%
%
\bibliographystyle{IEEEbib}
\ninept
\bibliography{MLSP21refs}

\begin{thebibliography}{10}

\bibitem{abesser2020review}
Jakob Abe{\ss}er,
\newblock ``A review of deep learning based methods for acoustic scene
  classification,''
\newblock {\em Applied Sciences}, vol. 10, no. 6, 2020.

\bibitem{DBLP:conf/ijcnn/AdavannePV18}
Sharath Adavanne, Archontis Politis, and Tuomas Virtanen,
\newblock ``Multichannel sound event detection using {3D} convolutional neural
  networks for learning inter-channel features,''
\newblock in {\em 2018 IEEE International Joint Conference on Neural Networks,
  ({IJCNN})}, Rio de Janeiro, Brazil, Jul. 2018, pp. 1--7.

\bibitem{DBLP:conf/icassp/AdavannePV17}
Sharath Adavanne, Pasi Pertil{\"{a}}, and Tuomas Virtanen,
\newblock ``Sound event detection using spatial features and convolutional
  recurrent neural network,''
\newblock in {\em 2017 {IEEE} International Conference on Acoustics, Speech and
  Signal Processing ({ICASSP})}, New Orleans, LA, USA, Mar. 2017, pp. 771--775.

\bibitem{DBLP:journals/taslp/WangC18a}
DeLiang Wang and Jitong Chen,
\newblock ``Supervised speech separation based on deep learning: An overview,''
\newblock {\em {IEEE} {ACM} Trans. Audio Speech Lang. Process.}, vol. 26, no.
  10, pp. 1702--1726, 2018.

\bibitem{DBLP:conf/asru/LuoHMCL19}
Yi~Luo, Cong Han, Nima Mesgarani, Enea Ceolini, and Shih{-}Chii Liu,
\newblock ``{FaSNet: L}ow-latency adaptive beamforming for multi-microphone
  audio processing,''
\newblock in {\em {IEEE} Automatic Speech Recognition and Understanding
  Workshop ({ASRU})}, Singapore, Dec. 2019, pp. 260--267.

\bibitem{DBLP:journals/eswa/GuimaraesNS20}
Heitor~R. Guimar{\~{a}}es, Hitoshi Nagano, and Diego~W. Silva,
\newblock ``Monaural speech enhancement through deep {Wave-U-Net},''
\newblock {\em Expert Syst. Appl.}, vol. 158, pp. 1--10, 2020.

\bibitem{DBLP:journals/corr/abs-1811-11307}
Craig Macartney and Tillman Weyde,
\newblock ``Improved speech enhancement with the {Wave-U-Net},''
\newblock {\em arXiv preprint: arXiv:1811.11307v1}, 2018.

\bibitem{DBLP:conf/eusipco/BoscaGPK20}
Am{\'{e}}lie Bosca, Alexandre Gu{\'{e}}rin, Laur{\'{e}}line Perotin, and Srdan
  Kitic,
\newblock ``Dilated {U-Net} based approach for multichannel speech enhancement
  from first-order {A}mbisonics recordings,''
\newblock in {\em 28th European Signal Processing Conference ({EUSIPCO})},
  Amsterdam, Netherlands, Jan. 2021, pp. 216--220.

\bibitem{DBLP:conf/icassp/HuangKHS14}
Po{-}Sen Huang, Minje Kim, Mark Hasegawa{-}Johnson, and Paris Smaragdis,
\newblock ``Deep learning for monaural speech separation,''
\newblock in {\em {IEEE} International Conference on Acoustics, Speech and
  Signal Processing ({ICASSP})}, Florence, Italy, May 2014, pp. 1562--1566.

\bibitem{DBLP:journals/speech/YanYWG20}
Xue Yan, Zhen Yang, Tingting Wang, and Haiyan Guo,
\newblock ``An iterative graph spectral subtraction method for speech
  enhancement,''
\newblock {\em Speech Commun.}, vol. 123, pp. 35--42, 2020.

\bibitem{DBLP:conf/interspeech/FanLTYW19}
Cunhang Fan, Bin Liu, Jianhua Tao, Jiangyan Yi, and Zhengqi Wen,
\newblock ``Discriminative learning for monaural speech separation using deep
  embedding features,''
\newblock in {\em 20th Annual Conference of the International Speech
  Communication Association ({INTERSPEECH})}, Gernot Kubin and Zdravko Kacic,
  Eds., Graz, Austria, Sep. 2019, pp. 4599--4603.

\bibitem{DBLP:journals/taslp/LuoM19}
Yi~Luo and Nima Mesgarani,
\newblock ``{Conv-TasNet}: {S}urpassing ideal time-frequency magnitude masking
  for speech separation,''
\newblock {\em {IEEE} {ACM} Trans. Audio Speech Lang. Process.}, vol. 27, no.
  8, pp. 1256--1266, 2019.

\bibitem{DBLP:journals/corr/abs-1905-01697}
Omolbanin Yazdanbakhsh and Scott Dick,
\newblock ``Multivariate time series classification using dilated convolutional
  neural network,''
\newblock {\em arXiv preprint: arXiv:1905.01697v1}, 2019.

\bibitem{DBLP:journals/taslp/PolitisMAHV21}
Archontis Politis, Annamaria Mesaros, Sharath Adavanne, Toni Heittola, and
  Tuomas Virtanen,
\newblock ``Overview and evaluation of sound event localization and detection
  in {DCASE} 2019,''
\newblock {\em {IEEE} {ACM} Trans. Audio Speech Lang. Process.}, vol. 29, pp.
  684--698, 2021.

\bibitem{DBLP:journals/corr/abs-2006-01919}
Archontis Politis, Sharath Adavanne, and Tuomas Virtanen,
\newblock ``A dataset of reverberant spatial sound scenes with moving sources
  for sound event localization and detection,''
\newblock {\em arXiv preprint: arXiv:2006.01919v2}, 2020.

\bibitem{DBLP:journals/jstsp/AdavannePNV19}
Sharath Adavanne, Archontis Politis, Joonas Nikunen, and Tuomas Virtanen,
\newblock ``Sound event localization and detection of overlapping sources using
  convolutional recurrent neural networks,''
\newblock {\em {IEEE} J. Sel. Top. Signal Process.}, vol. 13, no. 1, pp.
  34--48, 2019.

\bibitem{DBLP:conf/eusipco/GuirguisSGAY20}
Karim Guirguis, Christoph Schorn, Andre Guntoro, Sherif Abdulatif, and Bin
  Yang,
\newblock ``{SELD-TCN:} {S}ound event localization {\&} detection via temporal
  convolutional networks,''
\newblock in {\em 28th European Signal Processing Conference ({EUSIPCO})},
  Amsterdam, Netherlands, Jan. 2021, pp. 16--20.

\bibitem{chytas2019hierarchical}
Sotirios~Panagiotis Chytas and Gerasimos Potamianos,
\newblock ``Hierarchical detection of sound events and their localization using
  convolutional neural networks with adaptive thresholds,''
\newblock in {\em Proc. of the Detection and Classification of Acoustic Scenes
  and Events 2019 Workshop (DCASE)}, 2019, pp. 50--54.

\bibitem{DBLP:journals/corr/abs-1905-00268}
Yin Cao, Qiuqiang Kong, Turab Iqbal, Fengyan An, Wenwu Wang, and Mark~D.
  Plumbley,
\newblock ``Polyphonic sound event detection and localization using a two-stage
  strategy,''
\newblock {\em arXiv preprint: arXiv:1905.00268v4}, 2019.

\bibitem{DBLP:journals/corr/abs-1910-04388}
Luca Mazzon, Yuma Koizumi, Masahiro Yasuda, and Noboru Harada,
\newblock ``First order ambisonics domain spatial augmentation for {DNN}-based
  direction of arrival estimation,''
\newblock {\em arXiv preprint: arXiv:1910.04388v1}, 2019.

\bibitem{pratik2019sound}
Pranay Pratik, Wen~Jie Jee, Srikanth Nagisetty, Rohith Mars, and Chongsoon Lim,
\newblock ``Sound event localization and detection using {CRNN} architecture
  with mixup for model generalization,''
\newblock in {\em Proc. of the Detection and Classification of Acoustic Scenes
  and Events 2019 Workshop (DCASE)}, 2019, pp. 199--203.

\bibitem{farina2000simultaneous}
Angelo Farina,
\newblock ``Simultaneous measurement of impulse response and distortion with a
  swept-sine technique,''
\newblock in {\em 108th Convention of the Audio Engineering Society}, Paris,
  France, Feb. 2000.

\bibitem{DBLP:conf/icassp/PanayotovCPK15}
Vassil Panayotov, Guoguo Chen, Daniel Povey, and Sanjeev Khudanpur,
\newblock ``Librispeech: An {ASR} corpus based on public domain audio books,''
\newblock in {\em 2015 {IEEE} International Conference on Acoustics, Speech and
  Signal Processing ({ICASSP})}, South Brisbane, Queensland, Australia, Apr.
  2015, pp. 5206--5210.

\bibitem{DBLP:journals/corr/abs-2010-00475}
Eduardo Fonseca, Xavier Favory, Jordi Pons, Frederic Font, and Xavier Serra,
\newblock ``{FSD50K:} an open dataset of human-labeled sound events,''
\newblock {\em arXiv preprint: arXiv:2010.00475v1}, 2020.

\bibitem{DBLP:conf/eusipco/ChazanHHGG19}
Shlomo~E. Chazan, Hodaya Hammer, Gershon Hazan, Jacob Goldberger, and Sharon
  Gannot,
\newblock ``Multi-microphone speaker separation based on deep {DOA}
  estimation,''
\newblock in {\em 27th European Signal Processing Conference, ({EUSIPCO})}, {A}
  Coru{\~{n}}a, Spain, Sep. 2019, pp. 1--5.

\bibitem{DBLP:journals/corr/abs-2010-11566}
Ali Aroudi and Sebastian Braun,
\newblock ``{DBNET:} {DOA}-driven beamforming network for end-to-end farfield
  sound source separation,''
\newblock {\em arXiv preprint: arXiv:2010.11566v1}, 2020.

\bibitem{DBLP:conf/icassp/TaalHHJ10}
Cees~H. Taal, Richard~C. Hendriks, Richard Heusdens, and Jesper Jensen,
\newblock ``A short-time objective intelligibility measure for time-frequency
  weighted noisy speech,''
\newblock in {\em {IEEE} International Conference on Acoustics, Speech, and
  Signal Processing ({ICASSP})}, Dallas, Texas, {USA}, Mar. 2010, pp.
  4214--4217.

\bibitem{DBLP:conf/nips/BaevskiZMA20}
Alexei Baevski, Yuhao Zhou, Abdelrahman Mohamed, and Michael Auli,
\newblock ``wav2vec 2.0: {A} framework for self-supervised learning of speech
  representations,''
\newblock in {\em Advances in Neural Information Processing Systems (NeurIPS)},
  Dec. 2020.

\bibitem{DBLP:conf/waspaa/MesarosAPHV19}
Annamaria Mesaros, Sharath Adavanne, Archontis Politis, Toni Heittola, and
  Tuomas Virtanen,
\newblock ``Joint measurement of localization and detection of sound events,''
\newblock in {\em 2019 {IEEE} Workshop on Applications of Signal Processing to
  Audio and Acoustics ({WASPAA})}, New Paltz, NY, USA, Oct. 2019, pp. 333--337.

\end{thebibliography}
%
%
\end{document}